# Representing Contextualized Information in the NSDL

Carl Lagoze[1], Dean Krafft[1], Tim Cornwell[1], Dean Eckstrom[1],
Susan Jesuroga[2], Chris Wilper[1]

[1] Computing and Information Science, Cornell University, Ithaca, NY 14850 USA
{lagoze, dean, cornwell, eckstrom, cwilper}@cs.cornell.edu
http://www.cis.cornell.edu
[2] UCAR-NSDL, PO Box 3000, Boulder, CO 80307 USA
jesuroga@ucar.edu

**Abstract.** The NSDL (National Science Digital Library) is funded by the National Science Foundation to advance science and math education. The initial product was a metadata-based digital library providing search and access to distributed resources. Our recent work recognizes the importance of context – relations, metadata, annotations – for the pedagogical value of a digital library. This new architecture uses Fedora, a tool for representing complex content, data, metadata, web-based services, and semantic relationships, as the basis of an information network overlay (INO). The INO provides an extensible knowledge base for an expanding suite of digital library services.

## 1    Introduction

Libraries, traditional and digital, are by nature information rich environments - the organization, selection, and preservation of information are their raison d'etre. In pursuit of this purpose, libraries have focused on two areas: building a collection of all the *resources* that meet the library's selection criteria, and building a catalog of *metadata* that facilitates organization and discovery of those resources.

This is the approach that the NSDL (National Science Digital Library) Project took over its first three years of existence, when it focused mainly on the location and development of resources appropriate for Science, Technology, Engineering, and Mathematics education, and the creation of quality metadata about those resources. This focus was reflected in the technical infrastructure that harvested metadata from distributed providers, processed and stored that metadata, and made it available to digital library services such as search and preservation.

The value of an excellent collection of resources as a basis for library quality is undeniable. And, even after years of advances in automatic indexing, metadata remains important for a class of resources and applications. However, our three years of effort in the NSDL have revealed that collection building and metadata aggregation are necessary but not sufficient activities for building an information-rich digital library. In particular, our experience has led to two conclusions. First, the technical and organizational infrastructure to support harvesting, aggregation, and refinement of metadata is surprisingly human-intensive and expensive [15]. Second, in a world of increasingly powerful and ubiquitous search engines, digital libraries must distinguish themselves by providing more than simple search and access [16]. This is par-ticularly true in educationally-focused digital libraries where research shows the importance of interaction with information rather than simple intake.

Based on these conclusions, we have redirected our efforts over the past year towards building a technical infrastructure that supports a more refined definition of information richness. This definition includes, of course, collection size and integrity, and it accommodates the relevance of structured metadata. But it adds the notion of building *information context* around digital library resources. Our goal is to create a knowledge environment that supports aggregation of multiple types of structured and unstructured information related to library resources, the instantiation of multiple relationships among digital library resources, and participation of users in the creation of this context. We are creating an infrastructure that captures the wisdom of users [32], adding information from their usage patterns and collective experience to the formal resources and structured metadata we already collect.

Our technical infrastructure is based on the notion of an *information network overlay* [16] – a directed, typed graph that combines local and distributed information resources, web services, and their semantic relationships. We have implemented this infrastructure using Fedora [17], an architecture for representing complex objects and their relationships.

In this paper we describe the motivations for this architecture, present the information model that underlies it, and provide results from our year of implementation. We note for the reader that this is still a work in progress. The results we provide in this paper relate to the implementation and scaling issues in creating a rich information model. As our work progresses, we will report in future papers on the effectiveness of this architecture from the perspective of the user and evaluate whether it really does enable a richer and more useful digital library.

The organization of this paper is as follows. Section 2 describes related work and situates this work in the context of other digital library efforts. Section 3 summarizes the importance of information contextualization for educational digital libraries. Section 4 provides a brief background on the NSDL and establishes the application context in which this work occurs. Section 5 describes the information model of the information network overlay. Section 6 provides the results of our implementation experience. Finally, section 7 concludes the paper.

## 2 Related Work

The work described in this paper builds on a number of earlier and ongoing research and implementation projects that investigate the role of user annotations in information environments, the importance of inter-resource relationships, and the integration of web services with digital content. We believe that our work is distinguished from these other projects in two ways. First, it combines traditional digital library notions of resources and structured metadata with service-oriented architecture and semantic web technology, thereby representing the rich relationships among a variety of structured, unstructured, and semi-structured information. Second, it implements this rich information environment at relatively large scale (millions of resources), exercising a number of state-of-the-art technologies beyond their previous deployments.

Perhaps the most closely related work is the body of research on information annotation. Catherine Marshall has written extensively on this subject [20] in the digital library and hypertext context. A number of systems have been developed that implement annotation in digital libraries. For example, Roscheisen, Mogensen, and Winograd created a system call ComMenter [31] that allowed sharing of unstructured comments about on-line resources. The multi-valent document work at Berkeley provides the interface and infrastructure for arbitrary markup and annotation of digital documents, and storage and sharing of that markup [34]. The semantic web community has also examined annotation, with the Annotea project [13] being the most notable example.

The importance of annotation capabilities for education and scholarly digital libraries has been noted by many researchers including Wolfe [35]. The ScholOnto project [24] created a system for the publication and discussion/annotation of scholarly papers, arguing for the importance of informal information along-side established resources. Constantopoulus, et al. [8] examine the semantics of annotations in the SCHOLNET project, a EU-funded project to build a rich digital library environment supporting scholarship. Within the NSDL effort, there have been a number of projects that support annotations, most notably DLESE (Digital Library for Earth System Education) [1].

Annotations and their association with primary resources are one class of the variety of relationships that can be established among digital content. Ever since Vannevar Bush invented hypertext [5], researchers have been examining tools for inter-linking information. Faaborg and Lagoze [11] examined the notion of semantic browsing whereby users could establish personalized and sharable semantic relationships among existing web pages. Huynh, et al. [12] have recently done similar work in the Simile project.

There is also related work on resource linking specifically for pedagogic purposes within the educational research community. Unmil, et al. [33] describe Walden's Paths, a project that allows teachers to establish meta-structure over the web graph for creation of lesson plans and other learning materials. Recker, et al. have created another system called Instructional Architect [28], that similarly allows integration of on-line resources by teachers into educational units.

Finally, an important component of the work described here is the integration of content and web services. In many ways our digital library "philosophy" resembles that of the Web 2.0 philosophy [25]. Key components of this are the collection and integration of unique data, the participation of users in that data collection and formulation process, and the availability of the data environment as a web service that can be leveraged by value-add providers. Chad and Miller [6] extend Web 2.0 to something they call Library 2.0. We hope that our work demonstrates many of the principles they describe, notably the notion that Library 2.0 encourages a "culture of participation" and provides the interface to its accumulated information for innovative "mash-ups" that exploit library information in innovative ways.

## 3   The Need for Context and Reuse

Research shows that education-focused digital libraries (and digital libraries in general) need to support the full life cycle of information [19]. Reeves wrote "The real power of media and technology to improve education may only be realized when students actively use them as cognitive tools rather than simply perceive and interact with them as tutors or repositories of information." [30].

One requirement that appears frequently in the learning technology literature is the reuse of resources for the creation of new learning objects. This involves integrating and relating existing resources into a new learning context. A learning context has many dimensions including social and cultural factors; the learner's educational system; and the learner's abilities, preferences and prior knowledge [21].

Most digital libraries, including the NSDL, currently rely on forms of metadata to describe learning objects and enable discovery. Metadata standards abstract properties of learning objects, and abstraction can lead to instances where learning context is ignored or reduced to single dimensions [26]. Metadata is often focused on the technical aspects of description and cataloging, not on capturing the actual context of instructional use. Recker and Wiley write "a learning object is part of a complex web of social relations and values regarding learning and practice. We thus question whether such contextual and fluid notions can be represented and bundled up within one, unchanging metadata record." [29]

McCalla also argues that there is no way of guaranteeing that metadata captures the breadth and depth of content domains. He writes that, ideally, learning objects need to reflect "appropriateness" to address the differences between learners' needs. [22] In addition, questions remain as to whether these logical representations (e.g. metadata and vocabularies) created primarily for use by computer systems will make the most intuitive sense for learners [7].

Several approaches have been suggested to help supply the rich context for learning object creation and reuse. These include capturing opinions about learning objects and descriptions of how they are used [26]; recording the community of users from which the learning object is derived [29]; collecting teacher-created linkages to state education standards [28]; tracking and using student-generated search keywords [2]; and providing access to comments or reviews by other faculty and students [23].

We see that in order to provide an educationally-focused digital library, the information infrastructure must support flexible integration of information, ranging from highly structured metadata to unstructured comments and observations. It needs to nr dynamic, expanding both in the manner that a standard library collection expands, but also based on the collective experience and input of the user community.

## 4   A Suite of Contextualized NSDL Services

We are creating the infrastructure to meet notions of information richness outlined in the previous section. This work follows more than three years of work by the NSDL Core Infrastructure (CI) team, and has been described in a number of other papers

[14, 15]. Stated very briefly, this earlier work used OAI-PMH to populate a metadata repository (MR). This metadata was indexed by a CI-managed search service, which was accessible by users through a central portal at http://nsdl.org.

Our goal is to move beyond the search and access capabilities provided by the MR. The creation of the NSDL Data Repository (NDR), built on the architecture described in the next section, provides a platform for a number of exciting new NSDL applications focused directly on increasing user participation in the library. In addition to creating specific new capabilities for NSDL users, these applications all create context around resources that aids in discovery, selection and use. Four specific applications that exploit the infrastructure described in this paper are currently in various phases of development, testing, and deployment.

*Expert Voices* (EV) is a collaborative blogging system that fully integrates the resources and capabilities of the NDR. It allows subject matter experts to create realtime entries on critical STEM issues, while weaving into their presentation direct references to NSDL resources. These blog entries automatically become both resources in the NSDL library and annotations on all the referenced resources. EV supports Question/Answer discussions, resource recommendations and annotations, the provision of structured metadata about existing resources, and establishing relationships among existing resources in the NSDL, as well as between blog entries and resources.

*On Ramp* is a system for the distributed creation, editing, and dissemination of content from multiple users and groups in a variety of formats. Disseminations range from publications like the NSDL Annual Report to educational workshop materials to online presentations like the Homepage Highlights exhibit at NSDL.org's homepage. Resources created and released in OnRamp can become NDR content resources, and NDR resources and other content can be directly incorporated into On Ramp publications, creating new context and relationships within the NDR.

*Instructional Architect*, described by Recker [27], "… enables users (primarily teachers) to discover, select, and design instruction (e.g., lesson plans, study aids, homework) using online learning resources. ". Currently, IA supports searching the NSDL for resources and incorporating direct references to those resources into an IA project. The IA team is currently working with the NDR group to support both publication of IA projects as new NSDL resources and the direct capture the web of relationships created by an IA project in the NDR.

The *Content Alignment Tool* (CAT), currently in development by a team led by Anne Diekema and Elizabeth Liddy of Syracuse University, uses machine learning techniques to support the alignment of NSDL resources to state and national educational standards [10]. Initially (2Q2006), users will be able to use the tool to suggest appropriate educational standards for any resource they are viewing. Later versions of the system will allow experts and other users to provide feedback, incorporated into the NDR, on the appropriateness of the assignments. This tool, and the overall incorporation of educational standards relationships into the NDR, will allow NSDL users to search and browse the NSDL by "standards", starting either from a standard or from any relevant resource.

## 5   Design and Information Model

To provide the foundation for this rich array of user-visible services, we have implemented the NSDL Data Repository (NDR). The NDR implements all features of the pre-existing MR such as metadata harvesting, storage, and dissemination. However, it moves from the restrictive metadata-centric focus of the MR to a resource-centric model, which allows representation of rich relationships and context among digital library resources.

The NDR implements a data abstraction that we call an information network overlay (INO). Like other overlay networks [3] the INO instantiates a layer over another network, in this case the web graph.

Specifically, an INO is a directed graph. Nodes are identified via URIs and are packages of multiple streams of data. This data stream composition corresponds to compound object formats such as METS [18] and DIDL [4], allowing the creation of compound digital objects with multiple representations. The component data streams may be contained data or they may be surrogates (via URLs) to web-accessible content. This allows nodes to aggregate local and distributed content, for example the reuse of multiple primary resources into new learning objects. Web services may be associated with information units and their components, allowing service-mediated disseminations of the data aggregated in a digital object. This advances the reuse paradigm beyond simple aggregation, allowing, for example, a set of resources written in English to be refactored into a Spanish learning object though mediation by a translation service. Edges represent ontologically-typed relationships among the digital objects. The relationship ontology is extensible in the manner of OWL-based ontologies [9]. This allows the NDR to represent the variety of application-based relations described earlier such as collection membership, aggregation via reuse into a learning object, and correlation with one or more state standards. Nodes (digital objects) are polymorphic - they can have multiple types in the data model, where typing means the set of operations that can be performed on the digital object. In the digital library environment, this flexibility overcomes well-known dilemmas such as the data/metadata distinction, which conflicts with the reality that an individual object can be viewable through both of these type lenses.

The NDR is implemented within a Fedora repository. A complete description of Fedora is out-of-scope for this paper and the reader is directed to the up-to-date explanation at [17]. Each node in the INO corresponds to a Fedora digital object. Fedora provides all the functionality necessary for the INO including compound objects, aggregation of local and distributed content, web service linkages, and expression of semantic relationships. Fedora is implemented as a web service and includes fine-grained access control and a persistent storage layer.

Length constraints on this paper prohibit a full description of the information modeling in the NDR and the use of Fedora to accomplish this modeling. This modeling includes the design of Fedora digital objects to provide the different NDR object types – resources, agents, metadata, aggregations, and the like – and the relationships among these types for common use cases such as resource and metadata branding and resource annotation.

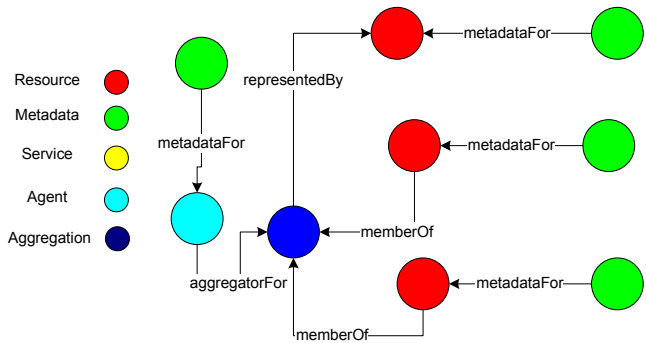

**Figure 1 - Modeling an aggregation**

However, an example shown in Figure 1 demonstrates how the NDR represents aggregation. Examples of aggregations include conventional collection/item membership, but also aggregations with other semantics such as membership of individual resources in a compound learning object or alignment of set of resources with a state educational standard.  Each node corresponds to a Fedora digital object, with the key at the left showing the type of the object.  The labels on the arcs document the type of the relationship.  As shown, "memberOf" arcs relate resources to one or more aggregations.  Aggregations can have arbitrary semantics, with the semantics documented by the resource that is the object of the "representedBy" arc.  For example, this resource may be a surrogate for a collection, or may represent a specific state standard.  Lastly, the person or organization responsible for the aggregation is represented by the agent that is the source of the "aggregatorFor" arc.

## 6   Results from Implementation of the NSDL Data Repository

Over the past year we have been designing, implementing, and loading data into the NDR.  The major implementation task was the creation and coding of an NDR-specific API for manipulation of information objects in the NDR data model – specific "types" of digital objects such as resources, metadata, agents, and the like and the required relationships among them.  Note that this API is distinct from the SOAP and REST API in Fedora that provides access to low-level digital object operations.  The NDR API consists of a set of higher level operations such as addResource, addMetadata, and setAggregationMembership.  Each of these higher level operations is a composition of low-level Fedora primitive operations. For example, the logical NDR operation addResource, which adds a new resource to the NDR, translates to a set of low-level Fedora operations including creating the digital object that corresponds to the resource, configuring its datastreams so that they match our model for the resource "type", and establishing the relationships between that resource and its collection digital object and to the metadata digital objects that describe it.

   We have implemented in Java an API layer that mediates all interaction with the NDR, by calling on the constituent set of low-level Fedora operations.  In addition to providing a relatively easy-to-use interface for services accessing the NDR, the API performs the vital task of ensuring that constraints of the data model are enforced. For example, the data model mandates that no metadata digital object should exist that does not have one (and only one) "metadataFor" relationship to a resource digital object.

We have used this API to bootstrap the production NDR with data from the pre-existing MR, thereby duplicating existing functionality in the new infrastructure. At the time of writing of this paper, this process is complete. The platform for our NDR production environment is a Dell 6850 server with dual 3Ghz Xeon processors, 32Gb of 400Mhz memory and 517Gb of SCSI RAID disk with 80MB/second sustained performance. This server is running 64-bit LINUX, for reasons outlined later. We note that the 2006 cost for this production server is about 22K USD.

The NDR has over 2.1 million digital objects – 882,000 of them matching metadata from the MR, 1.2 million of them representing NSDL resources, and several hundred representing other information objects – agents, services, etc., - in the NDR data model. The representation of the relationships among these objects (those defined by the NDR data model and those internal to the Fedora digital object representation) produces over 165 million RDF triples in the triple-store. We have found that ingest into the NDR takes about .7 seconds per object – making data load for this rich information environment a non-trivial task.

This bootstrapping process has been a learning process in scaling up semantically-rich information environments. In order to understand the results, it is necessary to distinguish three components: core Fedora, the triple-store it uses to represent and query inter-object relationships, and the Proai[1] component that supports OAI-PMH.

Core Fedora is a web service application built on top of a collection of file-system resident XML documents (one file for each digital object) and a relational database that caches fragments and transformations of those documents for performance. These XML documents are relatively small and stable, and at present we are using about 21 GBytes of disk space to store these files across 39,000 directories. We have not experienced any scaling problems nor do we foresee any with this core architecture. In fact, as we expected from our knowledge of the Fedora implementation, basic digital object access is not really dependent on the size of the Fedora repository. For example, our tests on dissemination performance show that requests for metadata formats that are stored literally in the NDR are about 69 ms. Requests for formats that are crosswalked from stored formats using an XSLT transform service take about 480 ms.

The more challenging aspect of our data loading and implementation work has involved the triple-store. Relationships among Fedora digital objects, and therefore among nodes in the NDR graph, are stored persistently as RDF/XML in a datastream in the digital object and are indexed as RDF triples in a triple-store, which provides query access to the relationship graph. In the case of the NDR, this provides query functionality such as "return all resources related to a state standard, a specific collection, or in an OAI set".

---

[1] http://www.fedora.info/download/2.1/userdocs/services/oaiprovider-service.html

Triple-store technology is relatively immature. Scaling it up to accomplish our initial data load has been especially challenging. As part of our implementation of the Fedora relationship architecture (known as the resource index), we experimented with scaling and performance of a number of tripe-store implementations. Our extensive tests comparing Sesame[2], Jena[3], and Kowari[4] are available online[5]. One particular target of our testing was the performance of complex queries that involve multiple graph node joins – these are the types of queries we issue to perform OAI-PMH List Records operations that select according to metadata format, set, and date range. We found that Jena would not scale over a few tens of thousands of triples with complex query times approaching 20 minutes for complex queries over .5 million triples. Sesame can be configured in both native storage mode or on top of mysql. We found that Sesame-mysql, like Jena, was unable to return large results sets, producing an out-of-memory error due to accumulating the entire result set in memory. Our remaining tests comparing Sesame native to Kowari showed that for a database of several million triples Kowari was faster by a factor of 2 for simple queries, and by a factor of over 9000 for complex queries.

Although the Kowari implementation proved capable under controlled tests of high performance and scalability, we encountered a number of hurdles along the path of our data load. The apparent reality is that neither Kowari nor any other triple-store has been pushed to this scale. Such scale revealed unpleasant and previously undiscovered bugs, such as a memory leak that took months of effort to verify and find[6]. Furthermore, we have found that the hardware requirements to run a large-scale semantic web application are non-trivial. Kowari uses memory mapped indexes, which are both disk and memory-intensive. Presently the Kowari-based resource index requires over 54 GB of virtual memory, which is significantly larger than the 5 GB addressable by standard 32-bit processors and operating systems (thus the configuration of our production server described earlier).

In order to understand our results on semantic queries to the NDR resource index (storing 165 million triples), it makes sense to divide these queries into two classes. The first class of queries is relatively simple, such as those issued by a user application seeking all resources correlated with a state standard or another accessing all members of a collection. We have found that query performance in this case is on the order of 25ms for the simplest examples (no transitive joins over the graph) to about 250 ms for examples with 2-3 joins. The second class of queries are those that populate the NDR OAI server, Proai, which is a part of the Fedora service framework. Proai is an advanced OAI server that supports any metadata format available through the Fedora repository via direct datastream transcription or service-mediated dissemination. It operates over a MySQL database that is populated via resource-index queries to Fedora (in batch after an initial load and incrementally over the lifespan of the Fedora repository). The resource-index queries to populate Proai are quite complex with semantics such as "list all Fedora disseminations representing OAI-records of a certain format, and get their associated properties and set membership information". Such a query takes about 1 hour, when issued in batch over the fully loaded repository, and the combination of queries to pre-load the Proai database after the batch NDR load takes about 1-2 days. We note, however, that this load is only performed once on initial load of the NDR and that incremental updates, as information is added to the NDR, are much quicker.

---

[2] http://www.openrdf.org/

[3] http://jena.sourceforge.net/

[4] http://www.kowari.org/

[5] http://tripletest.sourceforge.net/

[6] http://prototypo.blogspot.com/2005/09/kowari-memory-leak-found-and-fixed.html

Proai performance is quite impressive. Throughput on an OAI-PMH ListRecords request is about 900 records per/second, and we have been able to harvest all Dublin Core records from the NDR (to populate our search indexes) in about 3 hours.

Our results provide hardware guidelines for large Fedora implementations that use the resource index. We have found that they greatly benefit from a machine with large real memory, high-speed disks, and high-performance disk controllers. The Dual Xeon processors provide an excellent match for Fedora processing allowing uniform execution partitioning of core Fedora, the NDR API, Proai and MySql processing among the 4 hyper threaded CPU cores available. CPU clock rate is a minor performance factor compared with the overall memory and I/O performance of the chassis. As of this writing, machines with more than 32GB of memory remain rare. Within 18 months we anticipate that machines having 64GB will become commonly available.

## 7 Conclusions

We have described in this paper our initial work in implementing an advanced infrastructure to support an information-rich NSDL. This infrastructure supports the integration and reuse of local and distributed content, the integration of that content with web services, and the contextualization of that content within a semantic graph.

The work described in this paper has advanced the state-of-the-art in two areas. First, it involves the innovative use of Fedora to represent an information network overlay. This data structure combines local and distributed content management, service-oriented architecture, and semantic web technologies. At a time when digital libraries need to move beyond the search and access paradigm, the INO supports contextualized and participatory information environments. Second, this work pushes the envelope on scaling issues related to semantic web technologies. Although RDF and the semantic web have existed for over 8 years, large-scale implementations still need to be demonstrated. Our experience with scaling the NDR is instructive to a number of other projects looking to build on top of semantic web technologies.

The results in this paper demonstrate only the basic functionality of the NDR. The basic operations, however, are the building blocks for the applications described in Section 4. In future papers, we will describe our experience with these applications and the ability of the NDR to support them in a highly scaled manner.

## Acknowledgements

We thank the entire NSDL CI team for their contributions to this work. The authors acknowledge the contributions of the entire Fedora team, especially Sandy Payette. The work described here is based upon work supported by the National Science Foundation under Grants No. 0227648, 0227656, and 0227888. Support for Fedora is provided by the Andrew W. Mellon Foundation.